\documentclass[a4paper]{jpconf}
\usepackage{graphicx}
\usepackage{amsmath,amssymb,amsfonts}

\begin{document}
\title{Energy, angular momentum and pressure force distributions inside nucleons}

\author{C Lorc\'e}

\address{CPHT, CNRS, Ecole Polytechnique, Institut Polytechnique de Paris, Route de Saclay, 91128 Palaiseau, France}

\ead{cedric.lorce@polytechnique.edu}

\begin{abstract}
We review some of the recent developments regarding mass, angular momentum and pressure forces inside hadrons. These properties are all encoded in the energy-momentum tensor of the system, which is described at the non-perturbative level in terms of gravitational form factors. Similarly to electromagnetic form factors, Fourier transforms of gravitational form factors allow one to map out the distribution of the above mechanical properties in position space, providing a whole new way of studying in detail the internal structure of hadrons.
\end{abstract}

\section{Introduction}
For more than 60 years, elastic scattering on the nucleon provided constraints on the spatial distribution of electric charge and magnetization inside nucleons~\cite{Perdrisat:2006hj,Miller:2010nz}. More recently, other exclusive reactions like deeply virtual Compton scattering and deeply virtual meson production started to provide constraints on the so-called generalized parton distributions (GPDs), which generalize both ordinary parton distributions and electromagnetic form factors~\cite{Diehl:2003ny,Belitsky:2005qn,Kumericki:2016ehc}. Beside providing multidimensional pictures of the nucleon internal structure in both position and momentum spaces, GPDs were shown to be related to gravitational form factors (GFFs)~\cite{Ji:1996ek} and hence to provide an indirect access to the energy-momentum tensor (EMT) of the system. 

By analogy with the electromagnetic form factor, Polyakov defined the spatial distribution of various mechanical properties of the nucleon like mass, angular momentum and pressure forces, based on the Fourier transform of GFFs~\cite{Polyakov:2002yz,Polyakov:2018zvc}. This provided a new exciting way of looking at the internal structure of the nucleon.

We present in this contribution a quick overview of selected recent works on the EMT in QCD and the associated mechanical properties.

\section{Parametrization of the EMT for spin-$1$ targets}
Thanks to continuous and discrete space-time symmetries, the elastic matrix elements of the EMT can be expressed in terms of GFFs, which are non-perturbative Lorentz-invariant functions, multiplied by some Lorentz tensors sandwiched between onshell polarization tensors. In the case of spin-1 targets, parametrizations for the symmetric conserved EMT have been proposed in~\cite{Holstein:2006ud,Abidin:2008ku} and for the non-conserved one in~\cite{Taneja:2011sy}. In~\cite{Cosyn:2019aio} we found that a GFF has been overlooked in the latter case, and we further generalized the parametrization to the asymmetric non-conserved EMT $T^a_{\mu\nu}$ ($a=$ quark or gluon)
\begin{equation}\label{spin1EMT}
\begin{aligned}
 \langle p^\prime, \lambda^\prime| T^a_{\mu\nu}(0) | p, \lambda \rangle
  &=
  -2P_\mu P_\nu\left[(\epsilon^{\prime*} \epsilon)
  \mathcal{G}^a_1(t)
  -
  \frac{(\Delta\epsilon^{\prime*} )(\Delta\epsilon )}{2M^2}
  \mathcal{G}^a_2(t)\right]
   \\ &
  - \frac{1}{2}(\Delta_\mu \Delta_\nu - \Delta^2 g_{\mu\nu})
  \left[(\epsilon^{\prime*} \epsilon)
  \mathcal{G}^a_3(t) 
  -
  \frac{(\Delta\epsilon^{\prime*} )(\Delta\epsilon )}{2M^2}
  \mathcal{G}^a_4(t)\right]\\ &
  +
  P_{\{\mu}\left( \epsilon^{\prime*}_{\nu\}} (\Delta \epsilon)
  - \epsilon_{\nu\}} (\Delta \epsilon^{\prime*}) \right)
  \mathcal{G}^a_5(t)
   \\ &
  +
  \frac{1}{2} \left[
    \Delta_{\{\mu}\left( \epsilon^{\prime*}_{\nu\}} (\Delta\epsilon)
    + \epsilon_{\nu\}} (\Delta\epsilon^{\prime*}) \right) 
    - \epsilon_{\{\mu}^{\prime*}\epsilon_{\nu\}} \Delta^2
    - g_{\mu\nu}(\Delta\epsilon^{\prime*})(\Delta\epsilon)
    \right]
  \mathcal{G}^a_6(t)
   \\ &
  +\epsilon_{\{\mu}^{\prime*}\epsilon_{\nu\}} M^2  \mathcal{G}^a_7(t)
  + g_{\mu\nu} M^2\left[ (\epsilon'^*\epsilon) \mathcal{G}^a_8(t)
  - \frac{(\Delta\epsilon'^*)( \Delta\epsilon)}{2M^2}   \mathcal{G}^a_9(t)\right]
    \\ &
  +
  P_{[\mu}\left( \epsilon^{\prime*}_{\nu]} (\Delta \epsilon)
  - \epsilon_{\nu]} (\Delta \epsilon^{\prime*}) \right)
  \mathcal{G}^a_{10}(t)
  +
  \Delta_{[\mu}\left( \epsilon^{\prime*}_{\nu]} (\Delta \epsilon)
  + \epsilon_{\nu]} (\Delta \epsilon^{\prime*}) \right)
  \mathcal{G}^a_{11}(t),
\end{aligned}
\end{equation}
where $P=(p'+p)/2$, $\Delta=p'-p$, and $t=\Delta ^2$. $\epsilon$ and $\epsilon'^*$ are the standard initial and final polarization four-vectors, $M$ is the mass of the spin-1 target, $\{\quad \}$ denotes symmetrization and $[\quad]$ denotes antisymmetrization.

\section{Mass decomposition and balance equations}
Considering the properly normalized forward matrix elements of the EMT allows one to derive both the mass decomposition and the balance equations~\cite{Lorce:2017xzd,Lorce:2018egm}. For a spin-$1$ target, the matrix element~\eqref{spin1EMT} in the $\Delta\to 0$ limit reduces to~\cite{Cosyn:2019aio}
\begin{equation}\label{FL}
\langle p, \lambda^\prime | T^a_{\mu\nu}(0) | p, \lambda \rangle
= 2p_\mu p_\nu\left[\mathcal G^a_1(0)+\frac{1}{6}\mathcal G^a_7(0)\right]-2g_{\mu\nu}M^2\left[\frac{1}{2}\mathcal G^a_8(0)+\frac{1}{6}\mathcal G^a_7(0)\right]-\mathcal T_{\mu\nu} M^2\mathcal G^a_7(0).
\end{equation}
The first two Lorentz structures on the RHS are spin-independent and hence appear for all targets. The last structure did not appear for spin-$0$ and spin-$1/2$ targets because it involves the tensor polarization $\mathcal T_{\mu\nu}=-\frac{1}{3}(g_{\mu\nu}-\frac{p_\mu p_\nu}{M^2})-\text{Re}(\epsilon_\mu\epsilon^*_\nu)$. Contraction of this amplitude with $\frac{p^\mu p^\nu}{M^2}$ and $-\frac{1}{3}(g^{\mu\nu}-\frac{p^\mu p^\nu}{M^2})$ together with a global normalization factor $1/2M$ defines the partial proper energy and isotropic pressure-volume work, respectively
\begin{equation}
U_a=\left[\mathcal G^a_1(0)-\frac{1}{2}\mathcal G^a_8(0)\right]M,\qquad W_a=\left[\frac{1}{2}\mathcal G^a_8(0)+\frac{1}{6}\mathcal G^a_7(0)\right]M.
\end{equation}
The tensor polarization generates an anisotropy in the partial proper pressure-volume work
\begin{equation}
W^{\mu\nu}_a=\mathcal T^{\mu\nu}\left[-\frac{1}{2}\mathcal G^a_7(0)\right]M.
\end{equation}
Poincar\'e invariance forces the forward matrix element of the total EMT to assume the form
\begin{equation}
\langle p, \lambda^\prime | T_{\mu\nu}(0) | p, \lambda \rangle
= 2p_\mu p_\nu, 
\end{equation}
and therefore implies the following mass decomposition and balance equations
\begin{equation}
M=\sum_{a}U_a,\qquad\sum_{a} W_a=0,\qquad\sum_{a} W^{\mu\nu}_a=0.
\end{equation}

\section{Energy and pressure force distributions}
Since one cannot in general separate the center of mass motion from the internal motion in relativity, relativistic spatial distributions usually depend on the average four-momentum $P$. A necessary condition for a quasi-probabilistic interpretation is that the frame should be chosen so that no energy is transfered to the system $\Delta^0=\vec P\cdot\vec\Delta/P^0=0$. Working in the Breit frame $\vec P=\vec 0$ therefore allows one to define the three-dimensional spatial distribution of the EMT~\cite{Polyakov:2002yz,Lorce:2018egm,Lorce:2017wkb}
\begin{equation}\label{EMTdist}
\langle T^a_{\mu\nu}\rangle(\vec r)=\int\frac{\textrm{d}^3\Delta}{(2\pi)^3}\,e^{-i\vec\Delta\cdot\vec r}\,\frac{1}{2P^0}\langle \tfrac{\vec\Delta}{2}, \lambda^\prime |T^a_{\mu\nu}(0) | -\tfrac{\vec\Delta}{2}, \lambda \rangle
\end{equation}
with $P^0=\sqrt{M^2+\frac{\vec\Delta^2}{4M^2}}$. Integrating this distribution over all space, we recover the forward limit discussed in the previous section. Using the parametrization of the EMT for an unpolarized spin-$1/2$ target, we observed that the spatial distribution~\eqref{EMTdist} takes the same form as that of an anisotropic spherically symmetric medium~\cite{Lorce:2018egm}
\begin{equation}
\langle T^a_{\mu\nu}\rangle(\vec r)\sim \left[\varepsilon_a(r)+p_{t,a}(r)\right]u_\mu u_\nu- p_{t,a}(r)g_{\mu\nu}+\left[p_{r,a}(r)-p_{t,a}(r)\right]\frac{x_\mu x_\nu}{r^2}
\end{equation}
with $u^\mu=g^{0\mu}$, $x^\mu=(0,\vec r)$, and $r=|\vec r|$. $\varepsilon_a(r)$ is the energy density, $p_{r,a}(r)$ is the radial pressure, and $p_{t,a}(r)$ is the transverse pressure.
\begin{figure}[h]
\begin{center}
\begin{minipage}{38pc}
\includegraphics[width=18pc]{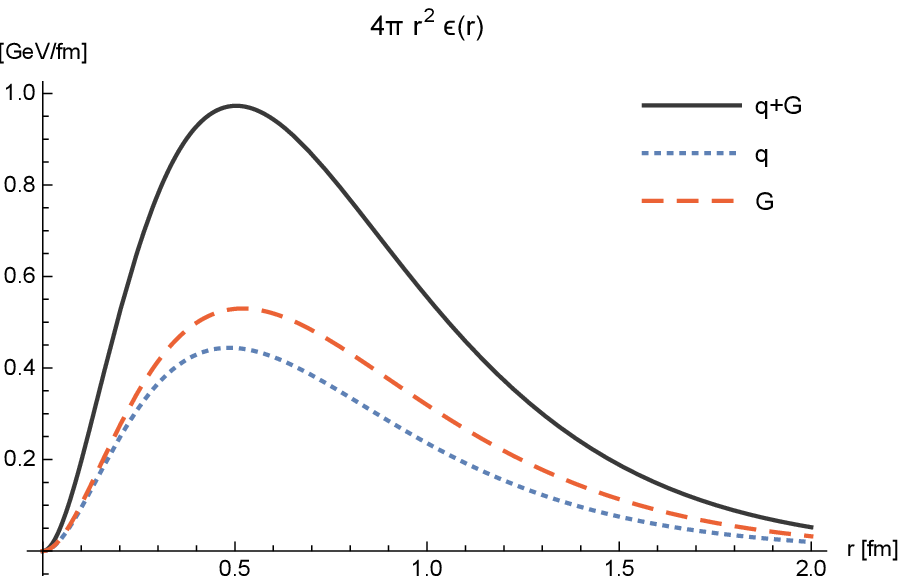}\hspace{2pc}\includegraphics[width=18pc]{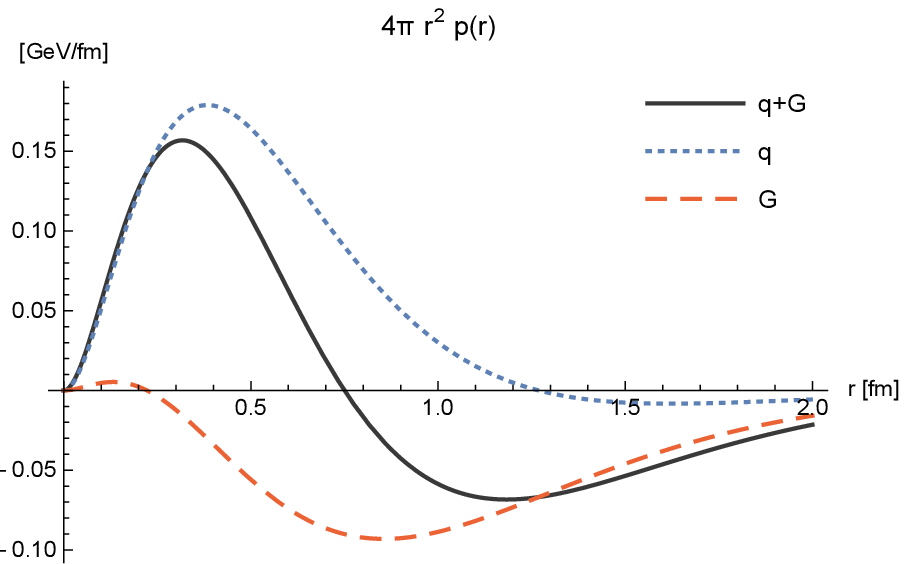}
\caption{\label{3Dplot}Distributions of energy $\varepsilon_a(r)$ and isotropic pressure $p_a(r)=\left[p_{r,a}(r)+2p_{t,a}(r)\right]/3$ in the nucleon using the multipole parametrization of~\cite{Lorce:2018egm}.}
\end{minipage}
\end{center}
\end{figure}
In Fig.~\ref{3Dplot} we show the distributions of energy density and isotropic pressure obtained within a multipole parametrization of the nucleon GFFs~\cite{Lorce:2018egm}. The energy density is always positive and is almost equally shared between quark and gluon contributions. Total isotropic pressure integrates to zero as expected from balance equations. Interestingly, the quark contribution is mostly repulsive and short range, while the gluon contribution is mostly attractive and long range. The difference of ranges between attractive and repulsive forces is responsible for pressure anisotropies $s_a(r)=p_{r,a}(r)-p_{t,a}(r)$. While pressure anisotropies are usually confined to very thin interfaces in ordinary media (and hence described in terms of a simple surface tension), in the case of the nucleon they turn out to be sizeable over a significant fraction of the volume attesting the relativistic nature of the system. 

Three-dimensional spatial distributions in the Breit frame are known to be plagued by relativistic corrections~\cite{Burkardt:2000za}. A convenient way out is to consider instead two-dimensional spatial distributions on the light front. In the case of the EMT, these have been introduced for the first time and discussed at length in~\cite{Lorce:2018egm,Lorce:2017wkb}.

\section{Angular momentum}
The spatial distribution of orbital angular momentum (OAM) can naturally be expressed in terms of the spatial distribution of momentum~\cite{Polyakov:2002yz,Lorce:2017wkb,Bakker:2004ib,Lorce:2011kd,Lorce:2011ni,Hatta:2011ku,Lorce:2012ce,Lorce:2015lna}
\begin{equation}\label{Lq}
\langle L^i_a\rangle (\vec r)=\epsilon^{ijk}\,r^j \langle T_a^{0k}\rangle(\vec r),
\end{equation}
see also the reviews~\cite{Leader:2013jra,Wakamatsu:2014zza,Liu:2015xha} for more details about the various types of OAM. Integrating this expression over all space and summing over all constituents gives the total OAM
\begin{equation}
\langle \vec L\rangle=\sum_a\int\textrm{d}^3r\,\langle \vec L_a\rangle (\vec r).
\end{equation}

If one works with the symmetric Belinfante form of the EMT, total OAM coincides with total angular momentum $\langle \vec J_\text{Bel}\rangle=\langle \vec L_\text{Bel}\rangle$. In the spin-$1/2$ case, Ji derived the following sum rule using the conservation of total angular momentum~\cite{Ji:1996ek}
\begin{equation}
\sum_a \left[A_a(0)+B_a(0)\right]=1,
\end{equation}
where $A_a(t)$ and $B_a(t)$ are GFFs entering the parametrization of the spin-$1/2$ EMT. Fundamentally, this sum rule is a consequence of Lorentz symmetry and can alternatively be derived using the boost operator~\cite{Lowdon:2017idv,Lorce:2018zpf}. A similar sum rule has been derived for spin-$1$ targets~\cite{Abidin:2008ku,Taneja:2011sy,Cosyn:2019aio} and more recently for any spin representation~\cite{Cotogno:2019xcl,Lorce:2019sbq}.

If one works with the asymmetric kinetic form of the EMT, an intrinsic spin contribution related to the antisymmetric part of the EMT has to be included to get the total angular momentum $\langle \vec J\rangle=\langle \vec L\rangle+\langle \vec S\rangle$. While both Belinfante and kinetic total angular momenta agree at the integrated level $\langle \vec J_\text{Bel}\rangle=\langle \vec J\rangle$, they differ at the distribution level
\begin{equation}
\langle \vec J_{\text{Bel},a}\rangle(\vec r)\neq \langle \vec L_a\rangle(\vec r)+\langle \vec S_a\rangle(\vec r).
\end{equation}
The difference noted $\langle \vec M_a\rangle(\vec r)$ is attributed to the superpotential used in the Belinfante procedure to eliminate the spin distribution by means of a reorganization of the momentum distribution. 

\begin{figure}[h]
\begin{center}
\begin{minipage}{38pc}
\includegraphics[width=18pc]{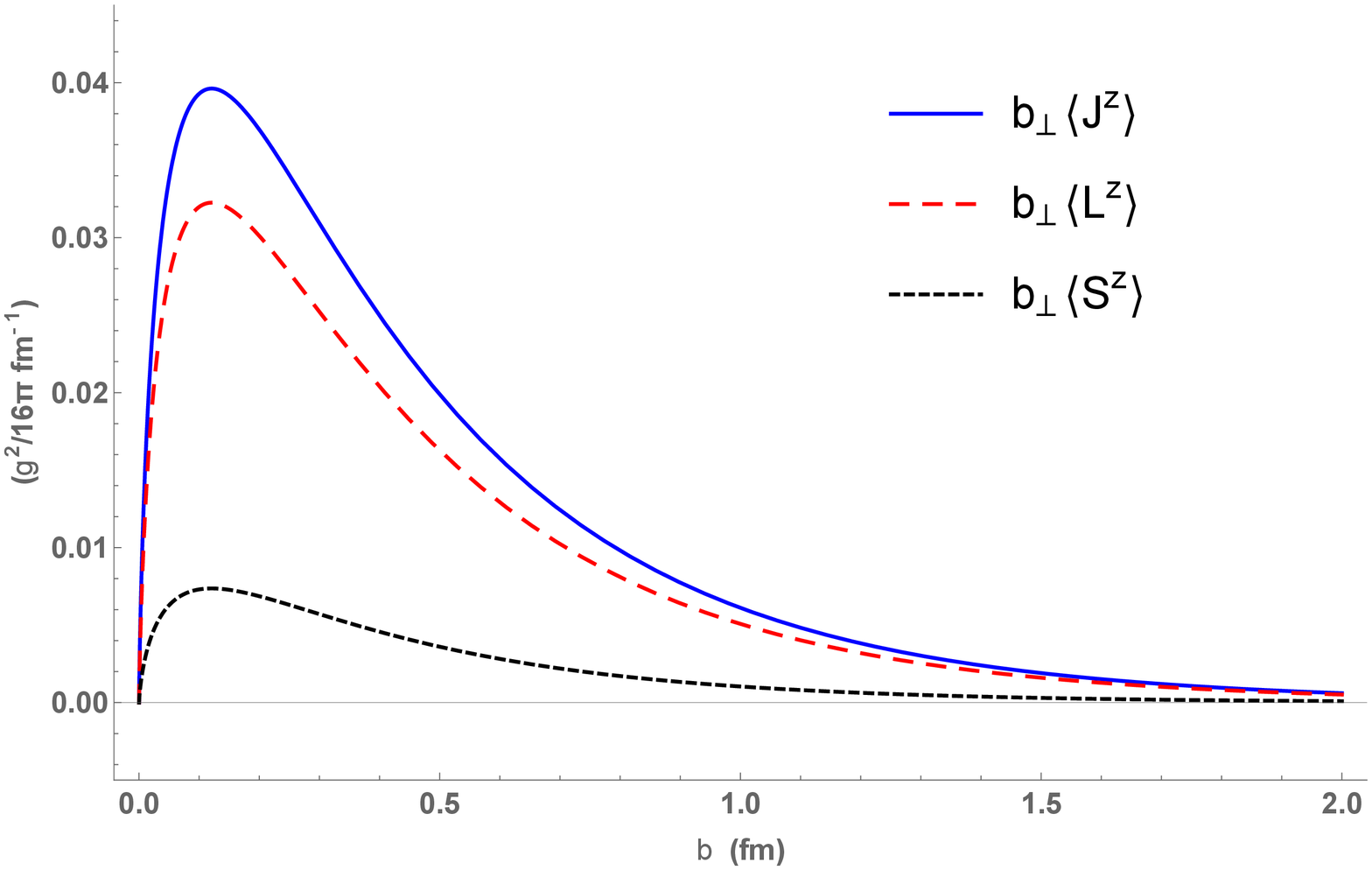}\hspace{2pc}\includegraphics[width=18pc]{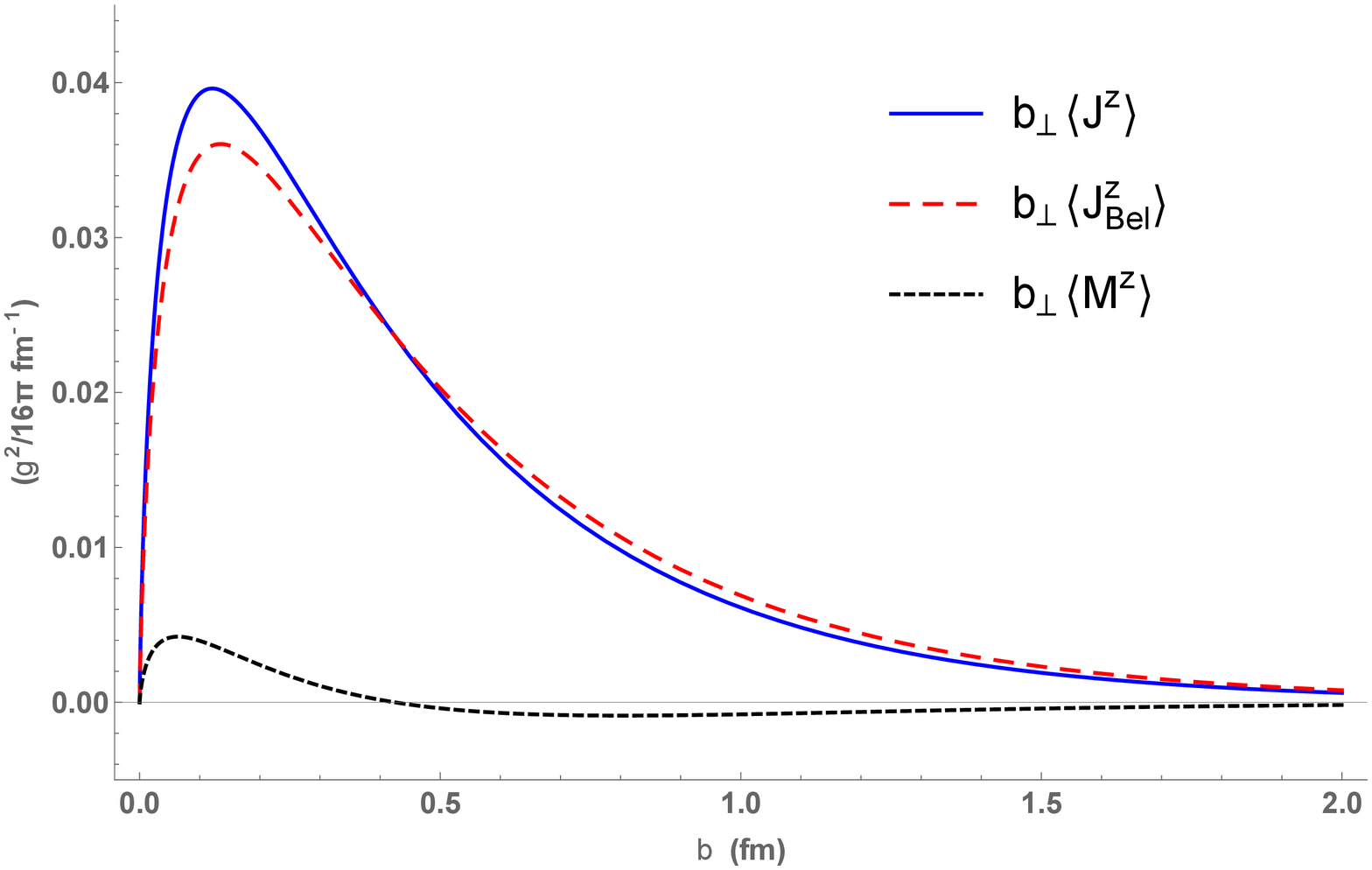}
\caption{\label{2Dplot} Light-front distributions of kinetic angular momentum (left) and comparison with the Belinfante angular momentum (right) using the scalar diquark model~\cite{Lorce:2017wkb}.}
\end{minipage}
\end{center}
\end{figure}

Like in the case of the EMT, two-dimensional distributions on the light front can be defined for the angular momentum~\cite{Lorce:2017wkb}. An illustration using the scalar diquark model is presented in Fig.~\ref{2Dplot}.

\section{Conclusion}
We presented a short overview of recent developments which illustrate the rich physics that can be addressed by studying the energy-momentum tensor. In particular, the question of the origin of the hadron mass and spin can now be studied at the level of spatial distribution, and stability conditions based on pressure forces might lead to new hints about the problem of confinement. For all these reasons, a lot of effort is being put in constraining gravitational form factors from both exclusive high-energy experiments and Lattice QCD.

\ack
The selected works presented in this contribution to the proceedings of INPC2019 have been supported by the Agence Nationale de la Recherche under the project No. ANR-16-CE31-0019 and ANR-18-ERC1-0002, the P2IO LabEx (ANR-10-LABX-0038) in the framework ``Investissements d'Avenir'' (ANR-11-IDEX-0003-01) managed by the Agence Nationale de la Recherche, the CEA-Enhanced Eurotalents Program co-funded by FP7 Marie Sklodowska-Curie COFUND Program (Grant agreement No. 600382), the U.S. Department of Energy, Office of Science, Office of Nuclear Physics, contract No. DE-AC02-06CH11357, and an LDRD initiative at Argonne National Laboratory under the project No. 2017-058-N0.

\end{document}